\documentclass[journal,twoside]{IEEEtran}

\usepackage{cite}
\usepackage[switch]{lineno}
\usepackage{amsmath,amssymb,amsfonts}
\usepackage{algorithmic}
\usepackage{graphicx}
\usepackage{textcomp}
\usepackage{framed,multirow}
\usepackage{booktabs}
\usepackage{multirow}
\usepackage{amsmath}
\usepackage{array}
\usepackage{verbatim}
\usepackage{amsmath}
\usepackage{amssymb}
\usepackage{xcolor}
\usepackage{multirow} 
\usepackage{tabularx}
\usepackage{colortbl}
\usepackage{xcolor}
\usepackage{array}
\usepackage{makecell}
\usepackage[ruled,linesnumbered]{algorithm2e}
\usepackage[colorlinks=true,
urlcolor=blue
]{hyperref}
\newcolumntype{C}[1]{>{\centering\arraybackslash}p{#1}}

\usepackage{amssymb}
\usepackage{latexsym}
\usepackage{float}
\usepackage{array}
\usepackage{enumitem}
\usepackage{caption}
\usepackage{tabularx}
\usepackage{booktabs}
\usepackage{pifont}
\usepackage{tabularx}
\usepackage{stfloats}
\usepackage{url}
\usepackage{xcolor}
\usepackage{hyperref}

\begin{document}

\title{BrainCSD: A Hierarchical Consistency-Driven MoE Foundation Model for Unified Connectome Synthesis and Multitask Brain Trait Prediction}

\author{Xiongri Shen, Jiaqi Wang, Yi Zhong, Zhenxi Song, Leilei Zhao, Liling Li, Yichen Wei, Lingyan Liang, Shuqiang Wang, \textit{Senior Member}, Baiying Lei, \textit{Senior Member}, Demao Deng, Zhiguo Zhang, \textit{Member}
\thanks{
This research was supported by the National Natural Science Foundation of China (Grants 62306089, 32361143787, 82102032), the China Postdoctoral Science Foundation (Grants 2023M730873, GZB20230960), the key Project of Basic Research of Shenzhen (NO: JCYJ20200109113603854), and the Guangxi Natural Science Foundation (Grant No. 2023GXNS-FBA026073), the Shenzhen Science and Technology Program(Grant No. RCBS20231211090800003) (Corresponding authors: Zhenxi Song and Zhiguo Zhang.)
}
\thanks{Xiongri Shen, Yi Zhong, Jiaqi Wang, Liling Li, and Leilei Zhao  are with the Department of Computer Science and Technology, Harbin Institute of Technology, Shenzhen, 518055, China (email: xiongrishen@stu.hit.edu.cn, zoey24@stu.hit.edu.cn, 23b951063@stu.hit.edu.cn, 24S151033@stu.hit.edu.cn, 24b951025@stu.hit.edu.cn ).}
\thanks{
Zhenxi Song and Zhiguo Zhang is with School of Intelligence Science and Engineering, College of Artificial Intelligence, Harbin Institute of Technology, Shenzhen, Guangdong, China(e-mail: songzhenxi@hit.edu.cn, zhiguozhang@hit.edu.cn).
}
\thanks{
Baiying Lei is with School of Biomedical Engineering, National-Regional Key Technology Engineering Laboratory for Medical Ultrasound, Guangdong Key Laboratory for Biomedical, Measurements and Ultrasound Imaging, Shenzhen University Medical School, Shenzhen University, Shenzhen, China (e-mail: leiby@szu.edu.cn).
}
\thanks{
Shuqiang Wang is with  the Shenzhen Institutes of
 Advanced Technology, Chinese Academy of Sciences, Shenzhen, China
 (e-mail: sq.wang@siat.ac.cn).
}
\thanks{
Demao Deng, Yichen Wei, and Lingyan Liangis with the Department of Radiology, The People’s Hospital of Guangxi Zhuang Autonomous Region, Guangxi Academy of Medical Sciences. Nanning, China (email: demaodeng@163.com, 316644690@qq.com, lianglingyan163@126.com).}
}


\maketitle

\begin{abstract}
Functional and structural connectivity (FC/SC) are key multimodal biomarkers for brain analysis, yet their clinical utility is hindered by costly acquisition, complex preprocessing, and frequent missing modalities. Existing foundation models either process single modalities or lack explicit mechanisms for cross-modal and cross-scale consistency. We propose BrainCSD, a hierarchical mixture-of-experts (MoE) foundation model that jointly synthesizes FC/SC biomarkers and supports downstream decoding tasks (diagnosis and prediction). BrainCSD features three neuroanatomically grounded components: (1) a ROI-specific MoE that aligns regional activations from canonical networks (e.g., DMN, FPN) with a global atlas via contrastive consistency; (2) a Encoding-Activation MOE that models dynamic cross-time/gradient dependencies in fMRI/dMRI; and (3) a network-aware refinement MoE that enforces structural priors and symmetry at individual and population levels. Evaluated on the datasets under complete and missing-modality settings, BrainCSD achieves SOTA results: 95.6\% accuracy for MCI vs. CN classification without FC, low synthesis error (FC RMSE: 0.038; SC RMSE: 0.006), brain age prediction (MAE: 4.04 years), and MMSE score estimation (MAE: 1.72 points). Code is available in \href{https://github.com/SXR3015/BrainCSD}{BrainCSD}

\end{abstract}
\vspace{-1.5mm}

\section{INTRODUCTION}
\label{sec:1}

\IEEEPARstart{E}{arly} and accurate diagnosis of neurodegenerative and neurodevelopmental disorders — including Alzheimer’s disease (AD) \cite{shang2023optimization,li2023developing,li2024pd,liu2024multi}, Parkinson’s disease (PD) \cite{cui2023adaptive,huang2024structural,pei2025transformer,jia2024assessing,zhang2025brainnet}, and Autism Spectrum Disorder (ASD) \cite{zhang2023detection,wang2024iFC,ren2023stratifying,li2025riemannian,rakshe2024autism,chen2025explainable} — is critical for timely intervention and improved long-term outcomes. Multi-modal neuroimaging, particularly resting-state functional MRI (fMRI) and diffusion MRI (dMRI) \cite{zhu2024jointly}, enables data-driven models. Recent work shows that \textit{connectivity biomarkers} — \textbf{functional connectivity (FC)} and \textbf{structural connectivity (SC)} — offer superior discriminative power \cite{cai2023individual}. These quantify synchronized neural activity (FC) and anatomical white-matter wiring (SC) between regions of interest (ROIs), providing a systems-level view of brain dysfunction \cite{ibrahim2021diagnostic}.

Yet two bottlenecks hinder clinical adoption. \textbf{First}, extracting FC/SC requires extensive preprocessing — slice-timing correction, motion realignment, co-registration, segmentation \cite{della2002empirical}, brain extraction, eddy correction, tractography \cite{zalesky2009dti} — taking \textit{months} per dataset. \textbf{Second}, real-world scans often lack fMRI/dMRI due to protocol limits or patient factors, leading to \textit{incomplete connectivity profiles} and performance collapse.

Recent foundation models have improved diagnostic generalizability. \textbf{BrainLM} \cite{carobrainlm} introduced masked self-supervision for intra-modality tasks (e.g., age prediction). \textbf{BrainSN} \cite{yang2025foundational} used ROI masking + Transformer for ASD/ADHD. \textbf{BrainMass} \cite{yang2024brainmass} built a biomarker-level model using atlas priors, covering AD/PD/ADHD. Yet none address missing-modality robustness — a critical gap for real-world deployment.

The Mixture-of-Experts (MoE) architecture has emerged as a key enabler for scaling foundation models \cite{rajbhandari2022deepspeed}, especially when combined with Transformers. Yet, current MoE research remains narrowly focused on improving gating mechanisms or router designs \cite{li2025uni} — largely ignoring \textit{domain-specific structural priors}. This is particularly limiting in diagnostics, where expert specialization should be guided by biologically meaningful units, not just data-driven routing signals.

\begin{figure*}[htbp]
    \centering
    \includegraphics[width=0.999\textwidth]{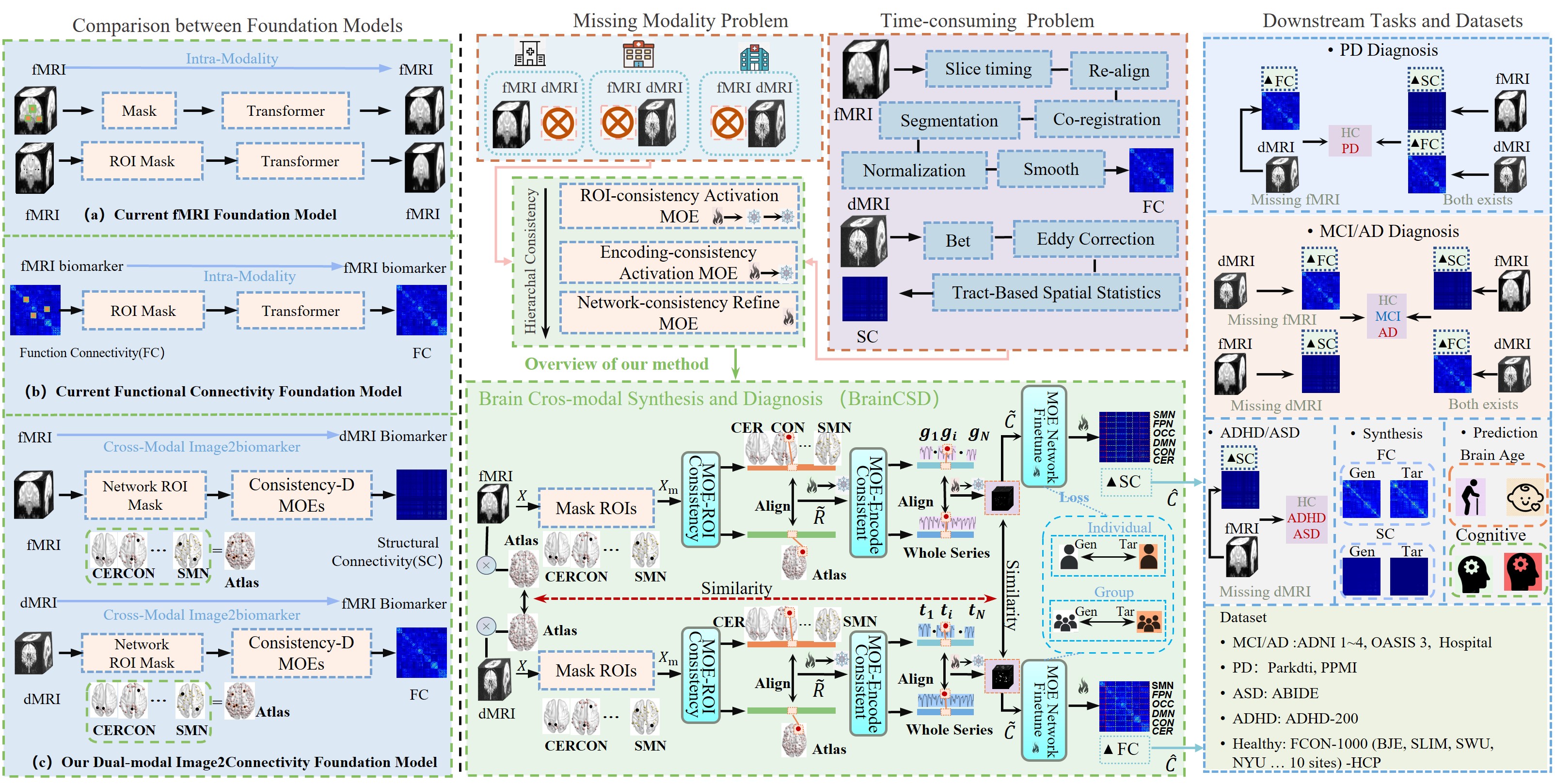} %
    \vspace{-3mm}
    \caption{ Hierarchical consistency-driven (consistency-D) MoE for connectome synthesis and prediction under missing modalities, via ROI (Network $\leftrightarrow$ Atlas) → Encoding  (time[\textbf{\textit{t}}]/gradient [\textbf{\textit{g}}] point $\leftrightarrow$ Whole series)→ Neuroanatomy-informed MOE.}
    \label{fig:method detail}
    \vspace{-4mm}
\end{figure*}

Although prior foundation models have embedded anatomical or functional priors into Transformer layers, no existing MoE explicitly structures expert routing around neuroanatomically defined brain networks — such as the Cerebellar Network (CER), Cingulo-Opercular Network (CON), Default Mode Network (DMN), Occipital Network (OCC), Frontoparietal Network (FPN), and Sensorimotor Network (SEN) — which are well-established as pathophysiologically critical in neurological and neurodevelopmental disorders \cite{zeng2024dynamic,ibrahim2021diagnostic,baggio2015cognitive}. This gap leaves MoE models biologically ungrounded and limits their interpretability, generalizability, and clinical utility — a limitation BrainCSD directly addresses through activation-consistency-driven, network-aligned expert routing.

We propose \textbf{BrainCSD} — \textbf{Brain} \textbf{C}onnectivity \textbf{S}ynthesis and \textbf{D}ecode — an \textbf{Hierarchical Consistency-Driven MOE Foundation Model} for joint medical image biomarker synthesis and decode. BrainCSD synthesizes FC/SC directly from raw fMRI/dMRI — bypassing preprocessing — and enables robust diagnosis even when one modality is missing.

Our \textbf{hierarchical consistency-driven MoE} enforces biological plausibility across three stages: (1) \textbf{ROI-consistency Activation MoE} — activates disease-relevant ROIs via atlas-MOEs
and network-MOEs alignment; (2) \textbf{Encoding Activation MoE} — treats each dual-modal neuro-imaging encoding-point as an expert to model dynamic FC/SC with temporal/gradient consistency between whole series MOEs; (3) \textbf{Network-Aware MoE Finetuning} — refines connectivity using individual and group structural constraints (symmetry, sparsity, para-graphics consistency).

\noindent\textbf{Contributions:}
\begin{enumerate}[leftmargin=*,noitemsep,topsep=0pt]
    \item We propose the hierarchical MoE foundation model that explicitly aligns expert routing with canonical brain networks (DMN, FPN, CON, etc.) through activation-consistency constraints across ROI, encoding, and population levels.
    
    \item We demonstrate that BrainCSD can synthesize diagnostically usable FC/SC from a single modality (e.g., dMRI-only), enabling MCI diagnosis with 95.6\% accuracy — outperforming real FC-based models in missing-modality settings.
    
    \item We validate robustness across 7K+ subjects, 22 sites, 15 tasks, 9 datasets,  and 5 disorders, showing $<$5\% performance variance across age, sex, and cognitive subgroups — a critical step toward clinical deployment.
\end{enumerate}

\begin{table*}[htbp]
\centering
\caption{Summary of Datasets Used in the Study }
\label{tab:dataset_summary}
\vspace{-2mm}
\resizebox{\linewidth}{!}{%
\begin{tabular}{@{}ccccccccc@{}}
\toprule
\textbf{Stage} & \textbf{Dataset} & \textbf{Sites} & \textbf{Modality} & \textbf{Participants} & \textbf{F/M} & \textbf{Age} & \textbf{MMSE/MoCA} & \textbf{Disease (nums)} \\ \midrule
\multirow{9}{*}{\shortstack{ROI/Encode \\ activation} } & ADNI 1-4 & ADNI 1-4 & fMRI-dMRI & 894 & 430/464 & 64.48 $\pm$ 5.73 & 29.55 $\pm$ 0.72 & HC(468)/MCI(370)/AD(56) \\
 & OASIS-3 & OASIS-3 & fMRI-dMRI & 1809 & 905/904 & 68.62 $\pm$ 4.73 & 28.90 $\pm$ 1.09 & HC(1315)/MCI(229)/AD(265) \\
 & Hospital & Hospital & fMRI-dMRI & 254 & 114/140 & 64.70 $\pm$ 6.47 & 29.00 $\pm$ 1.41 & HC(77)/MCI(177) \\
 & FCON-1000 & \begin{tabular}[c]{@{}l@{}}BIE, SLIM, SWU, HNU, BNU, \\ IPCAS, MRN, NKI, NYU, \\ SWU, XHCUMS\end{tabular} & fMRI-dMRI & 1705 & 853/852 & 24.30 $\pm$ 2.37 & / & HC(1705) \\
 & Park-dti & Park-dti & dMRI & 53 & 25/28 & 64.80 $\pm$ 7.44 & 28.10 $\pm$ 1.63 & HC(26)/PD(27) \\
 & PPMI & PPMI & fMRI-dMRI & 138 & 54/94 & 64.75 $\pm$ 8.78 & / & HC(18)/PD(120) \\
 & ABIDE & ABIDE & fMRI & 1073 & 149/924 & 16.60 $\pm$ 8.08 & / & HC(558)/ASD(515) \\
 & HCP & HCP & fMRI-dMRI & 1207 & 656/551 & 27.5 $\pm$ 2.94 & / & HC(1207) \\
 & ADHD-200 & ADHD-200 & fMRI & 194 & 49/145 & 11.4 $\pm$ 8.38 & / & HC(27)/167 \\ \midrule
Finetune-FC & \begin{tabular}[c]{@{}l@{}}ADNI, FCON-1000, \\ Hospital\end{tabular} & \begin{tabular}[c]{@{}l@{}}ADNI, FCON-1000, \\ Hospital\end{tabular} & fMRI-dMRI & 1277 & / & / & / & / \\ \midrule
Finetune-SC & ADNI & ADNI & fMRI-dMRI & 269 & / & / & / & / \\ \bottomrule
\end{tabular}%
}
\vspace{-4mm}
\end{table*}

\begin{table}[htbp]
\centering
\scriptsize
\setlength{\tabcolsep}{1.8pt}
\caption{Overview of downstream tasks and datasets.}
\vspace{-2mm}
\label{tab:downstream_tasks}
\begin{tabular}{@{}
    >{\raggedright\arraybackslash}p{1.3cm}  
    c                                       
    >{\raggedright\arraybackslash}p{3.2cm}  
    c                                       
    c                                       
@{}}
\toprule
\textbf{Scenario} & \textbf{Task} & \textbf{Dataset} & \textbf{Mod.} & \textbf{Subj.} \\
\midrule
\multicolumn{5}{@{}l}{\textbf{Diagnosis}} \\
\cmidrule(r){1-5}
\multirow{2}{1.8cm}[0pt]{Missing fMRI}  
 & MCI  & ADNI 1-4 & dMRI & 269 \\
 & PD   & Park-dti & dMRI & 53 \\
\addlinespace[1pt]
\multirow{3}{1.3cm}[0pt]{Missing dMRI}
 & MCI  & ADNI 1-4, Hospital, FCON-1000 & fMRI & 1277 \\
 & ASD  & ABIDE & fMRI & 1073 \\
 & ADHD & ADHD-200 & fMRI & 194 \\
\addlinespace[1pt]
\multirow{2}{1.3cm}[0pt]{Dual-modal Missing }
 & AD   & ADNI 1-4, OASIS-3, Hospital, HCP, FCON-1000 & f+D & 4934 \\
 & PD   & PPMI & f+D & 138 \\
\midrule
\multicolumn{5}{@{}l}{\textbf{Prediction}} \\
\cmidrule(r){1-5}
\multirow{2}{1.3cm}[0pt]{Dual Modal  Missing}
 & Age (w/ sub)  & ADNI 1-4, OASIS-3, Hospital, PPMI, HCP, FCON-1000 & f+D & 4934 \\
 & MMSE (w/ sub) & ADNI 1-4, OASIS-3, Hospital & f+D & 2957 \\
\midrule
\multicolumn{5}{@{}l}{\textbf{Synthesis}} \\
\cmidrule(r){1-5}
\multirow{2}{1.3cm}[0pt]{Missing dMRI}
 & FC (w/ sub) & ADNI 1-4, Hospital, FCON-1000 & fMRI & 1277 \\
\addlinespace[1pt]
\multirow{1}{1.8cm}{Missing fMRI}  
 & SC (w/ sub) & ADNI 1-4 & dMRI & 269 \\
\bottomrule
\end{tabular}

\vspace{1pt}
\footnotesize
\textbf{Note:} ADHD = Attention Deficit Hyperactivity Disorder; MMSE = Mini-Mental State Examination; f+D = fMRI + dMRI; (w/ sub) = with subgroup stratification.
\end{table}

\section{Related Work}
 \noindent\textbf{Connectivity Biomarkers for Diagnosis.}
Functional (FC) and structural connectivity (SC) biomarkers offer superior generalizability and neurobiological interpretability over raw imaging features~\cite{shang2023optimization,li2023developing,cui2023adaptive}. Disruptions in canonical networks (e.g., DMN, FPN) correlate strongly with disorders like AD, PD, and ASD~\cite{ren2023stratifying,li2025riemannian,zhang2023detection}, capturing systems-level pathology that enhances cross-dataset robustness.

\noindent\textbf{Limitations of Neuroimaging Foundation Models.}
Current foundation models (e.g., BrainLM~\cite{carobrainlm}, BrainSN~\cite{yang2025foundational}, BrainMASS~\cite{yang2024brainmass}) fall into \textit{image-level} or \textit{biomarker-level} categories. While biomarker-level models show stronger diagnostic transferability, they do not explicitly model robustness to missing modalities or eliminate dependency on slow, preprocessing-heavy pipelines, two critical requirements for real-world deployment that BrainCSD addresses by directly synthesizing FC/SC from uni-modality inputs.

\noindent\textbf{MoE with Neuroscientific Priors.}
MoE architectures excel in scalable representation learning~\cite{rajbhandari2022deepspeed,li2025uni}, yet generic routing mechanisms (e.g., top-$k$, attention-gating) often overlook domain-specific structure. As BrainMASS demonstrates, embedding neuroanatomical priors (e.g., atlas-based positional embeddings) improves performance. BrainCSD advances this by explicitly aligning MoE expert routing with canonical brain networks (DMN, FPN, CON, etc.) and temporal/gradient dynamics — ensuring that each expert specializes in a biologically meaningful subsystem.

\noindent\textbf{Synthesis Without Diagnostic Validation.}
Existing FC/SC synthesis methods, based on GANs, diffusion models, or VAEs~\cite{chen2025joint,zuo2024u,zhao2025diffusion,guan2025spatio,yuan2024remind} — primarily optimize statistical or visual fidelity. Few validate whether synthesized biomarkers preserve diagnostic signal, and even fewer evaluate under missing-modality settings. BrainCSD bridges this gap by jointly optimizing synthesis quality \textit{and} downstream tasks across complete and incomplete data regimes.

\section{Methods}
\label{sec:method}

We present \textbf{BrainCSD}, a MoE framework for synthesizing functional and structural connectomes (FC/SC) from resting-state fMRI and diffusion MRI (dMRI). As illustrated in Fig.~\ref{fig:method detail}, BrainCSD comprises three synergistic stages: (1) \textit{ROI Activation MoE} for spatial region selection grounded in neuroanatomical priors, (2) \textit{Encode Dynamics MoE} for modeling encoding-varying (time/gradient) brain states with a fixed set of encode experts, and (3) \textit{Network-Aware MoE Finetuning} for refining connectivity matrices with structural symmetry and network constraints. This hierarchical design enables precise, interpretable, and biologically plausible connectome synthesis.

\begin{figure*}[htbp]
    \centering
    \includegraphics[width=0.99\textwidth]{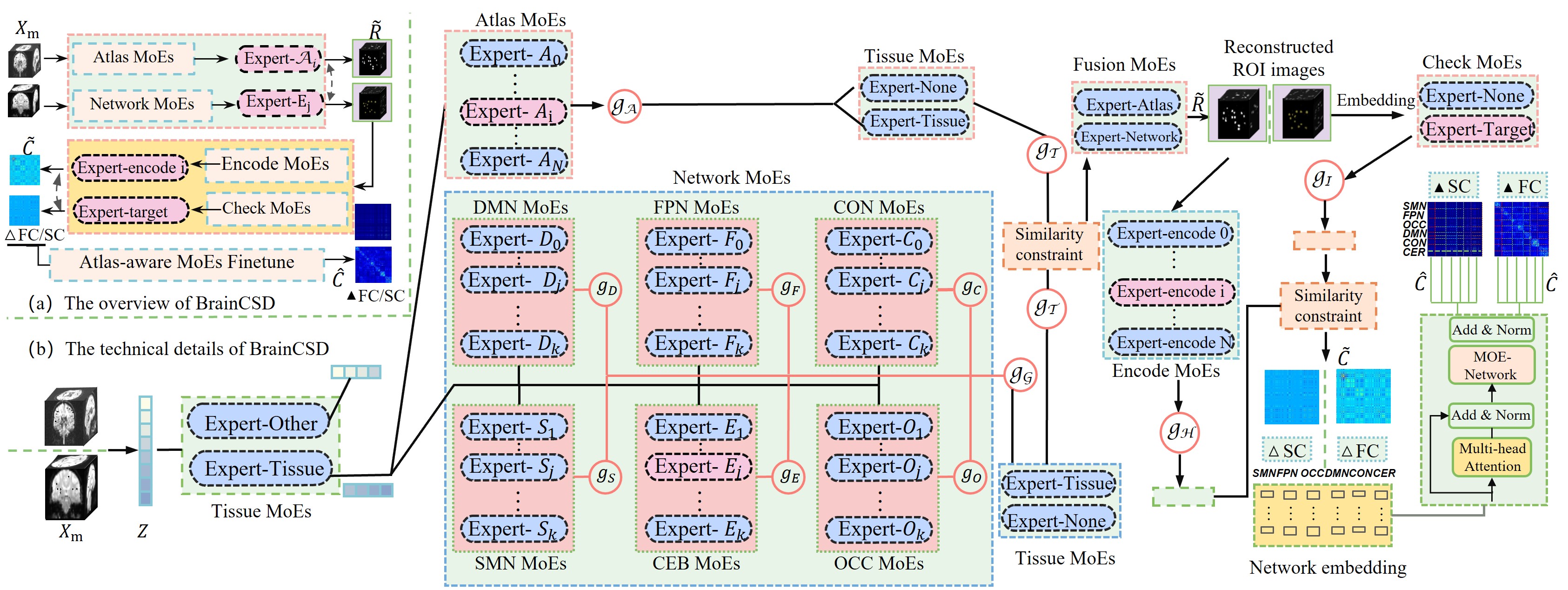} %
    \vspace{-2mm}
    \caption{ BrainCSD:  ROI → Encoding → Network refinement for hierarchical activation-consistent connectome synthesis}
    \label{fig:method detail}
    \vspace{-4mm}
\end{figure*}

\subsection{ROI Activation MoE}
\label{subsec:roi_moe}

\noindent \textbf{Motivation.} Neuroimaging data contains abundant non-informative background voxels and anatomically irrelevant regions. To focus on disease-diagnostic signals, we design a hierarchical MoE structure that first isolates brain tissue, then activates canonical networks (e.g., DMN, FPN), and finally enforces consistency between network-wise and whole-atlas selections via contrastive learning.

\vspace{0.5em}

\begin{figure}[htbp]
    \centering
        \vspace{-2mm}
    \includegraphics[width=0.45\textwidth]{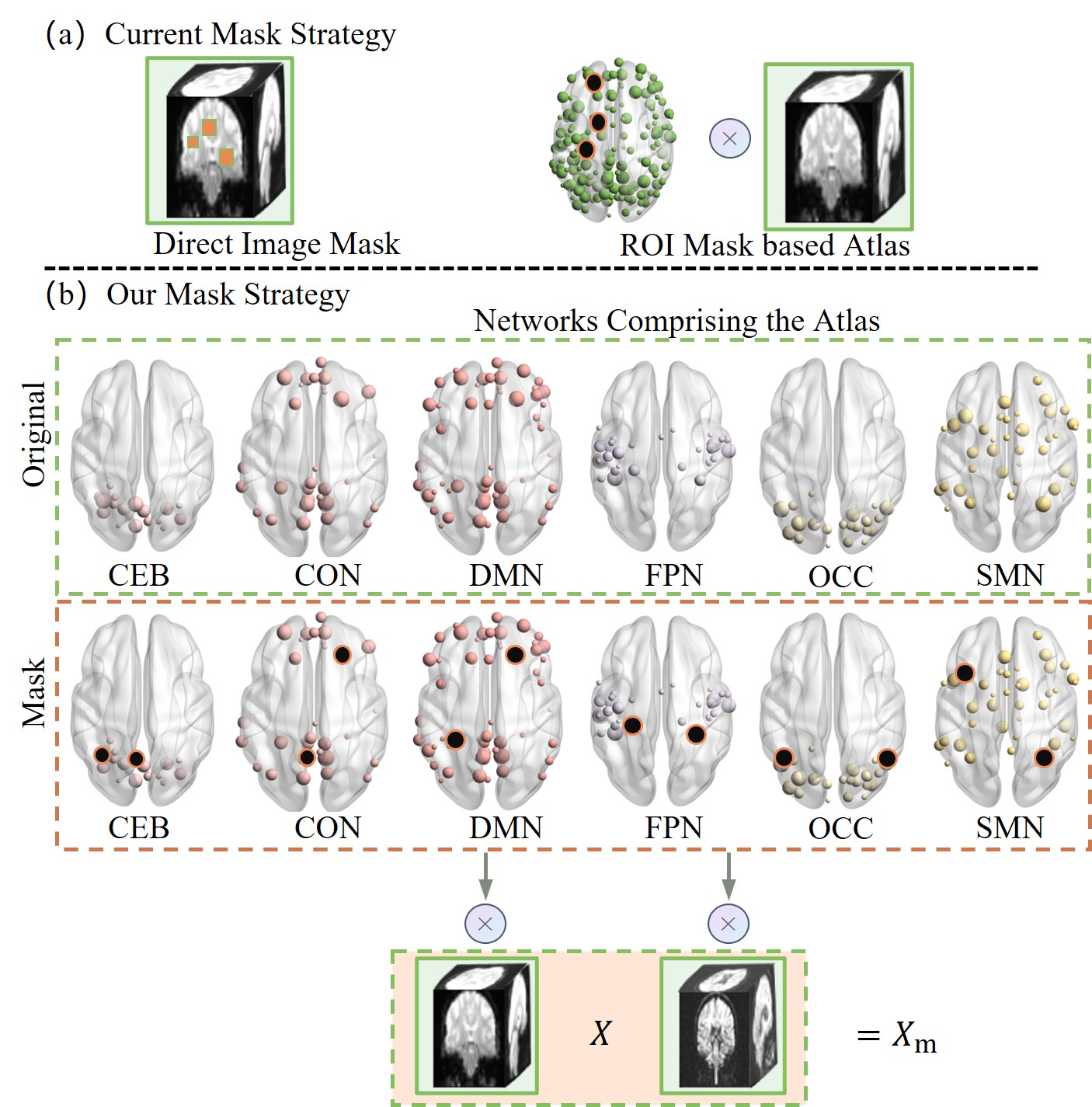}
        \vspace{-1mm}
    \caption{\textbf{Network-Level Masking Strategy.} Regions from six canonical networks are selectively masked during training to enforce region-specific feature learning.}
    \label{fig:mask strategy}
    \vspace{-4mm}
\end{figure}

\noindent \textbf{Network-Level Masking.}
Let $\mathbf{X}_f \in \mathbb{R}^{T \times V}$ and $\mathbf{X}_d \in \mathbb{R}^{G \times V}$ denote the fMRI encoding series and dMRI-derived gradient encoding seires, respectively, where $T/G$ is the number of time/gradient points and $V$ is the number of ROIs. Based on a standard brain atlas partitioned into six canonical networks — Cerebellar (CER), Cingulo-Opercular (CON), Default Mode (DMN), Occipital (OCC), Frontoparietal (FPN), and Somatomotor (SEN) — we define binary masks $\mathbb{M}_n \in \{0,1\}^V$ for each network $n$. The masked input is as:
\begin{equation}
    \mathbf{X}_m = \mathbf{X} \odot \left( \sum_{n=1}^6 \mathbb{M}_n \right),
\end{equation}
where $\odot$ denotes element-wise multiplication, and $\mathbf{X}$ represents either $\mathbf{X}_f$ or $\mathbf{X}_d$.

\vspace{0.5em}
\noindent \textbf{Tissue-Level MoE.}
To suppress non-brain background, we apply a Transformer encoder followed by a Tissue MoE $\mathcal{T} = \{ \mathcal{T}_1, \dots, \mathcal{T}_K \}$ with gating function $g_{\mathcal{T}}(\mathbf{Z}) = \mathrm{softmax}(\mathbf{W}_g \mathbf{Z})$. The embedding is:
\begin{equation}
    \mathbf{Z}_t = \sum_{i \in \mathrm{Top}\text{-}1(g_{\mathcal{T}}(\mathbf{Z}))} \left[ g_{\mathcal{T}}(\mathbf{Z})_i \cdot \mathcal{T}_i(\mathbf{Z}) \right],
\end{equation}
where $\mathbf{Z} = \mathrm{Transformer}(\mathbf{X}_m)$ is the initial embedded representation.

\vspace{0.5em}
\noindent \textbf{Network-Level MoE.}
For each neuro-anatomical network $\mathcal{N} \in \{ \mathcal{D}, \mathcal{F}, \mathcal{C}, \mathcal{S}, \mathcal{E}, \mathcal{O} \}$ (corresponding to DMN, FPN, etc.), we define a network-specific MoE $\mathcal{N}^r = \{ \mathcal{N}^r_1, \dots, \mathcal{N}^r_k \}$ with gating $g_{\mathcal{N}^r}$. The activated representation per network is:
\begin{equation}
    \mathbf{Z}_{\mathcal{N}^r} = \sum_{j \in \mathrm{Top}\text{-}1(g_{\mathcal{N}^r}(\mathbf{Z}_t))} \left[ g_{\mathcal{N}^r}(\mathbf{Z}_t)_j \cdot \mathcal{N}^r_j(\mathbf{Z}_t) \right].
\end{equation}

\vspace{-1mm}
\noindent \textbf{Global Consistency via Contrastive Learning.}
To align network-wise and atlas-wise activations, we introduce a global router $\mathcal{G}$ and an atlas-level MoE $\mathcal{A}$. Let $\mathbf{Z}_G = \mathcal{G}(\mathrm{Concat}(\{ \mathbf{Z}_{\mathcal{N}^r} \}))$ and $\mathbf{Z}_A = \mathcal{A}(\mathbf{Z}_t)$. We enforce consistency by applying tissue filtering again:
\begin{align}
    \mathbf{Z}_{tG} &= \sum_{l \in \mathrm{Top}\text{-}1(g_{\mathcal{T}}(\mathbf{Z}_G))} \left[ g_{\mathcal{T}}(\mathbf{Z}_G)_l \cdot \mathcal{T}_l(\mathbf{Z}_G) \right], \\
    \mathbf{Z}_{tA} &= \sum_{o \in \mathrm{Top}\text{-}1(g_{\mathcal{T}}(\mathbf{Z}_A))} \left[ g_{\mathcal{T}}(\mathbf{Z}_A)_o \cdot \mathcal{T}_o(\mathbf{Z}_A) \right].
\end{align}
A contrastive loss aligns these representations within a batch of $K$ subjects:
\begin{equation}
    \mathcal{L}_{\mathrm{cont}} = -\log \frac{
        \exp \left( s(\mathbf{Z}_{tA}, \mathbf{Z}_{tG}) / \tau \right)
    }{
        \sum_{k=1}^{K} \exp \left( s(\mathbf{Z}_{tA}, \mathbf{Z}_{tG}^{(k)}) / \tau \right)
    },
    \label{eq:constrastive_learning}
\end{equation}
where $s(\mathbf{u}, \mathbf{v}) = \frac{\mathbf{u}^\top \mathbf{v}}{\|\mathbf{u}\| \|\mathbf{v}\|}$ is cosine similarity and $\tau$ is a temperature hyperparameter.

\vspace{0.5em}
\noindent \textbf{Reconstruction to Imaging Space.}
The fused $\mathbf{Z}_f = \mathrm{Concat}(\mathbf{Z}_G, \mathbf{Z}_A)$ is decoded by a full-capacity MoE $\mathcal{R}$  (i.e., all experts contribute) to reconstruct ROI encoding series. The output is reshaped and positional encoding is removed to yield $\mathbf{R} \in \mathbb{R}^{E \times V}$:
\begin{equation}
 \widetilde{\mathbf{R}} = \mathrm{Reshape}\left( \mathrm{MLP}\left( \mathbf{Z}_f \right) \right) \in \mathbb{R}^{E \times V},
\end{equation}
where $E=T/V$. The final ROI response is reconstructed by removing positional encoding:
\begin{equation}
     \widetilde{\mathbf{R}}  = \mathbf{Z}_f - \mathbf{E}_{\mathrm{pos}}.
\end{equation}

\subsection{Encoding Activation MoE}
\label{subsec:time_moe}

\noindent \textbf{Motivation.} The connectivity is inherently dynamic — static ROI representations fail to capture encode evolution critical for accurate FC/SC estimation. We model encode dynamics (time axis in fMRI and gradient axis in dMRI) via a MoE with $K$ encode experts, each capturing a distinct phase of brain state evolution.

\vspace{0.5em}
\noindent Given the ROI response $ \widetilde{\mathbf{R}}  \in \mathbb{R}^{E \times V}$, E denotes encoding-vary point, we first embed it via Transformer 
to obtain $\mathbf{Z}_{ \widetilde{\mathbf{R}}} \in \mathbb{R}^{E \times D}$. A Encoding Activation MoE $\mathcal{H} = \{ \mathcal{H}_1, \dots, \mathcal{H}_T \}$ with gating $g_{\mathcal{H}}$ computes:
\vspace{-2mm}
\begin{equation}
\mathbf{Z}_E = \sum_{k=1}^{K} g_{\mathcal{H}}(\mathbf{Z}_{ \widetilde{\mathbf{R}}} )_k \cdot \mathcal{H}_k(\mathbf{Z}_{ \widetilde{\mathbf{R}}} )
\end{equation}
Then a check MOE $ \mathcal{I}$ with  gating $g_{\mathcal{I}}$ is used to embed the whose series:
\begin{equation}
    \mathbf{Z}_I = \sum_{p \in \mathrm{Top}\text{-}1(g_{\mathcal{I}}(\mathbf{Z}_{ \widetilde{\mathbf{R}}))}} \left[ g_{\mathcal{I}}(\mathbf{Z}_{ \widetilde{\mathbf{R}}} ) \cdot \mathcal{I}_p(\mathbf{Z}_{ \widetilde{\mathbf{R}}} ) \right],
\end{equation}
\vspace{-1mm}
To ensure encode stability, we apply a symmetric encode consistency loss:
\vspace{-1mm}
\begin{equation}
    \mathcal{L}_{\mathrm{en}} = \| \mathbf{Z}_E - \mathbf{Z}_I \|_F^2.
\end{equation}
The encoding-consistent ROI response $\mathbf{X}_T = \mathbf{Z}_E - \mathbf{E}_{\mathrm{pos}}$ is used to compute preliminary FC/SC matrices:
\vspace{-1mm}
\begin{equation}
    \widetilde{\mathbf{C}} = \mathrm{Corr}(\mathbf{X}_T),
\end{equation}
where $\mathrm{Corr}(\cdot)$ computes Pearson correlation for FC or streamline counts for SC.

\subsection{Network-Aware MoE Finetuning}
\label{subsec:finetune}

\noindent \textbf{Motivation.} Raw synthesized connectivity matrices often lack neurobiological plausibility — e.g., broken symmetry, unrealistic sparsity, or misaligned network boundaries. We introduce a finetuning stage that refines $\widetilde{\mathbf{C}}$ using network-constrained MoEs and structural priors. We vectorized $\widetilde{\mathbf{C}} \in \mathbb{R}^{V \times V}$ into $c \in \mathbb{R}^{V^2}$ and apply a linear projection before Transformer.

\vspace{0.5em}
\noindent Let $\mathbf{Z}_C = \mathrm{Transformer}(\widetilde{\mathbf{C}})$. For each network $\mathcal{N}^f \in \{ \mathcal{D}^f, \mathcal{F}^f, \dots, \mathcal{O}^f \}$, we apply a finetuning MoE:
\begin{equation}
    \mathbf{Z}_{\mathcal{N}^f} = \sum_{j \in \mathrm{Top}\text{-}1(g_{\mathcal{N}^f}(\mathbf{Z}_C))} \left[ g_{\mathcal{N}^f}(\mathbf{Z}_C)_j \cdot \mathcal{N}^f_j(\mathbf{Z}_C) \right].
\end{equation}
The refined connectivity $\hat{\mathbf{C}}$ is obtained by inverse embedding and enforces symmetry:
\vspace{-2mm}
\begin{equation}
    \hat{\mathbf{C}} = \mathrm{MLP}(\mathrm{Concat}(\{ \mathbf{Z}_{\mathcal{N}^f} \})) - \mathbf{E}_{\mathrm{pos}}, \quad
    \hat{\mathbf{C}} \leftarrow \frac{\hat{\mathbf{C}} + \hat{\mathbf{C}}^\top}{2}.
\end{equation}
\vspace{-2mm}
\begin{figure}[t]
    \centering
    \includegraphics[width=0.5\textwidth]{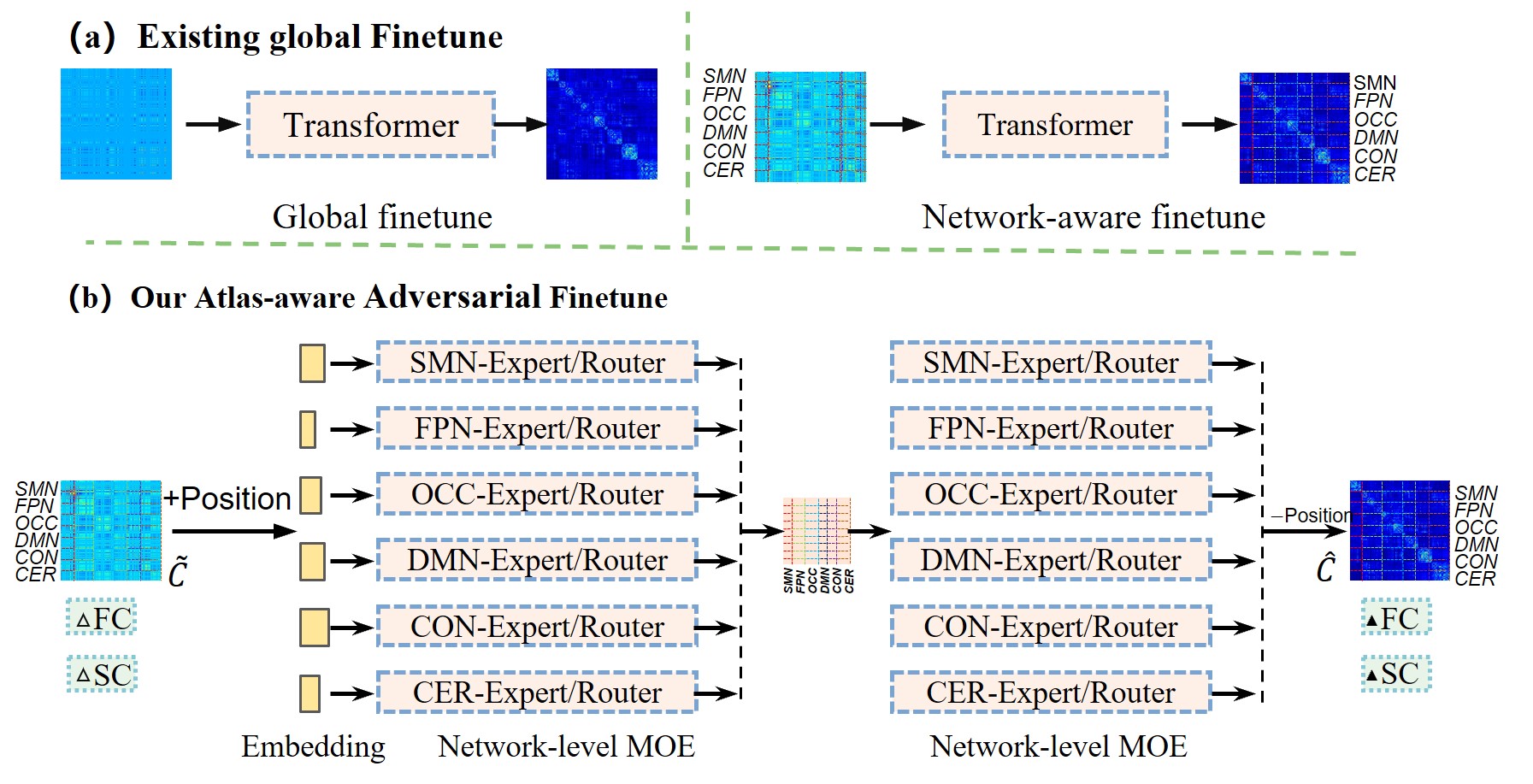}
    \vspace{-5mm}
    \caption{\textbf{Network-Aware Finetuning.} Each network is refined by a dedicated MoE expert, ensuring anatomical and diagnostic consistency.}
    \label{fig:atlas moe finetune}
    \vspace{-5mm}
\end{figure}

\subsection{Loss Functions}
\label{subsec:loss}

\begin{figure}[htbp]
    \centering
    \vspace{-3mm}
    \includegraphics[width=0.45\textwidth]{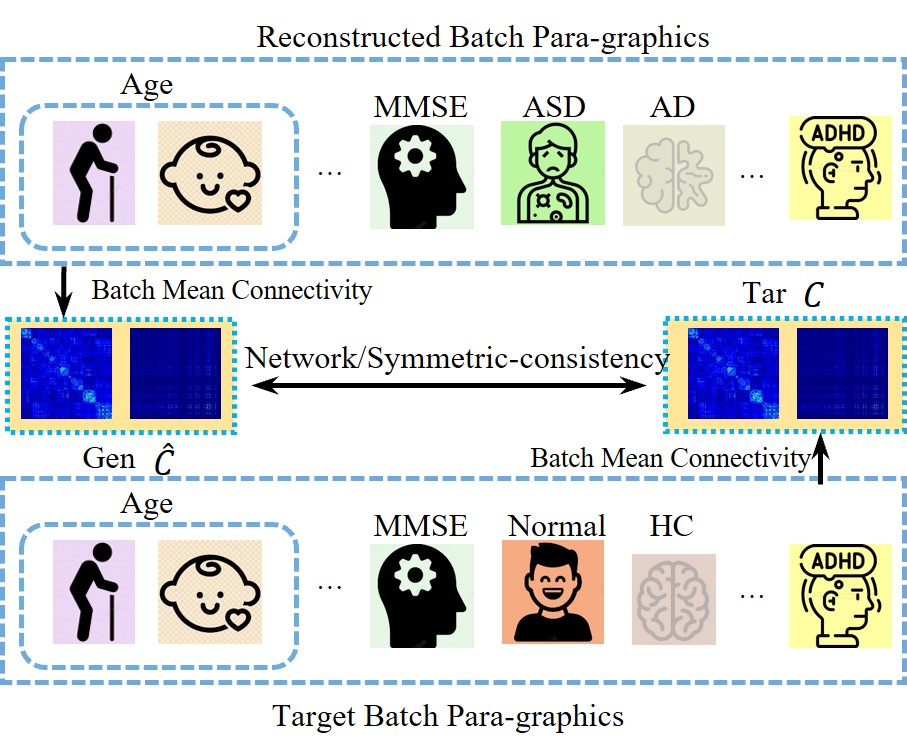}
    \vspace{-2mm}
    \caption{\textbf{Group-Level Consistency.} Connectivity features are regularized to be consistent within clinical/demographic subgroups, enhancing diagnostic utility.}
    \label{fig:group loss}
        \vspace{-3mm}
\end{figure}

\noindent \textbf{Reconstruction Loss.} At ROI and Encoding stages, we use mean squared error (MSE) and contrastive loss to reconstruct ROIs end encoding seiries, the $R$ is generated by apply the atlas to $X$, $i$ denotes f/d modality:
  \vspace{-1mm}

   \vspace{-4mm}
\begin{equation}
    \mathcal{L}_{\mathrm{cont\_e}} = \| \mathbf{R} - \hat{\mathbf{R}} \|_F^2  +
    \left\| 
        c(\mathbf{R}_1, \mathbf{R}_2) - c(\hat{\mathbf{R}}_1, \hat{\mathbf{R}}_2)
    \right\|_F^2,
\end{equation}
where $c$ is the contrastive loss function defined as:
\begin{equation}
    c(\mathbf{u}, \mathbf{v}) = -\log \frac{\exp(s(\mathbf{u}, \mathbf{v}) / \tau)}{\sum_{k=1}^{K} \exp(s(\mathbf{u}, \mathbf{v}) / \tau)},
\end{equation}
with $s(\mathbf{u}, \mathbf{v}) = \frac{\mathbf{u}^\top \mathbf{v}}{\|\mathbf{u}\| \|\mathbf{v}\|}$ being the cosine similarity and $\tau$ being a temperature hyperparameter.

\begin{figure*}[htbp]
\centering
\includegraphics[width=0.9\textwidth]{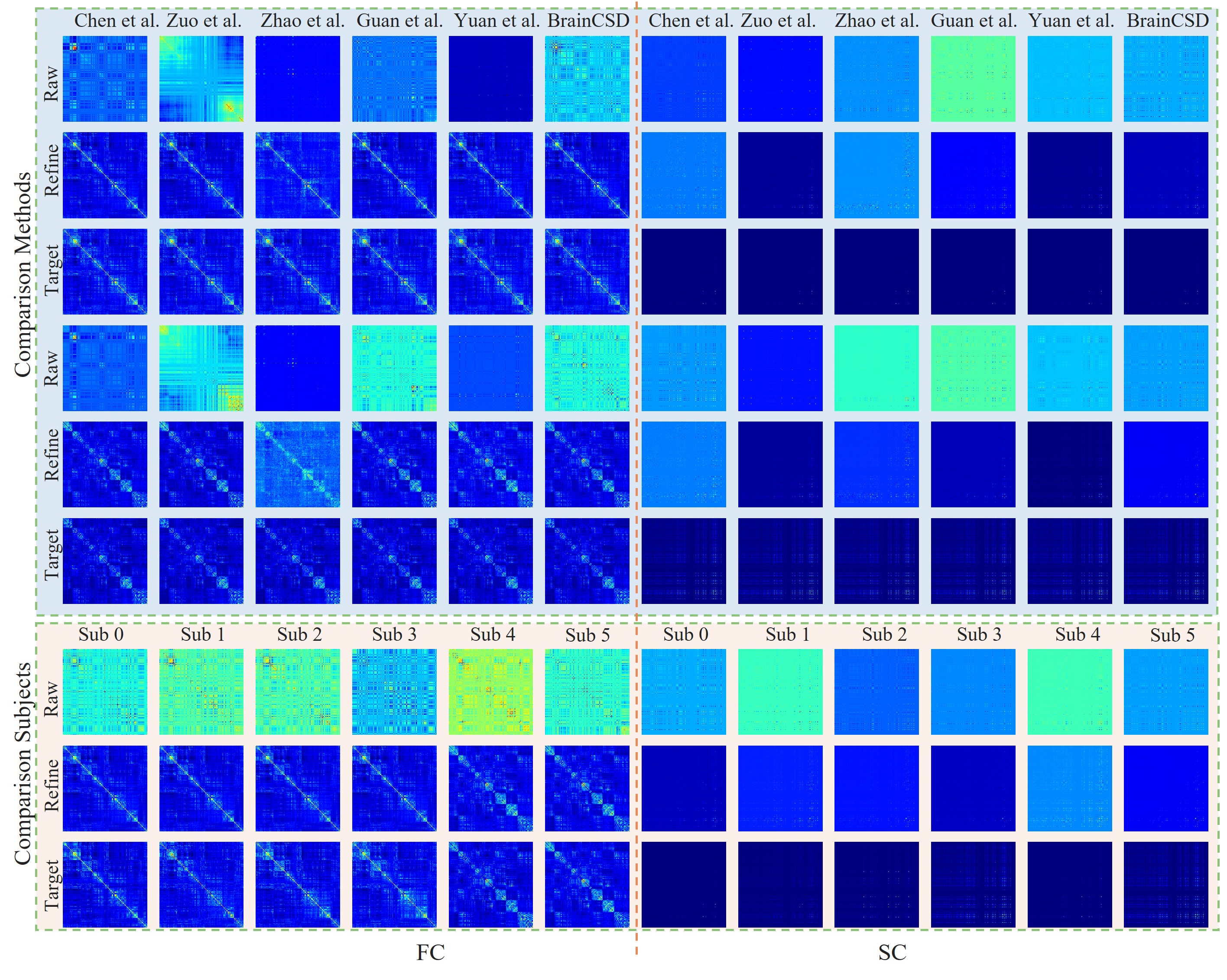}
\vspace{-4mm}
\caption{Visual comparison of FC/SC synthesis across methods. BrainCSD preserves network topology and inter-regional correlations, while baselines exhibit blurring or anatomical misalignment.}
\label{fig: syntheis comparision}
\vspace{-2mm}
\end{figure*}

\begin{table*}[htbp]
\centering
\scriptsize
\setlength{\tabcolsep}{2pt}
\caption{Synthesis quality (mean $\pm$ std) for FC/SC. ``+'' = raw output from method; ``--'' = refined.}
\label{tab:synthesis_combined}
\vspace{-2mm}
\begin{tabular}{@{}l c *{4}{c} *{4}{c}@{}}
\toprule
 & & \multicolumn{4}{c}{\textbf{FC}} & \multicolumn{4}{c}{\textbf{SC}} \\
\cmidrule(lr){3-6} \cmidrule(lr){7-10}
Method & Data & MSE & SSIM & RMSE & MAE & MSE & SSIM & RMSE & MAE \\
\midrule

\multicolumn{10}{@{}l}{\textbf{Domain Methods}} \\
Chen et al.~\cite{chen2025joint} & + & 0.126$\pm$0.022 & 0.059$\pm$0.036 & 0.353$\pm$0.034 & 0.213$\pm$0.022 & 0.001$\pm$0.000 & 0.411$\pm$0.148 & 0.036$\pm$0.006 & 0.009$\pm$0.001 \\
Chen et al.~\cite{chen2025joint} & -- & 0.002$\pm$0.002 & 0.789$\pm$0.126 & 0.040$\pm$0.013 & 0.029$\pm$0.012 & 0.001$\pm$0.000 & 0.586$\pm$0.149 & 0.022$\pm$0.008 & 0.005$\pm$0.001 \\
Zuo et al.~\cite{zuo2024u} & + & 0.096$\pm$0.013 & 0.113$\pm$0.029 & 0.309$\pm$0.022 & 0.228$\pm$0.023 & 0.001$\pm$0.000 & 0.395$\pm$0.152 & 0.033$\pm$0.005 & 0.009$\pm$0.001 \\
Zuo et al.~\cite{zuo2024u} & -- & 0.002$\pm$0.002 & 0.790$\pm$0.126 & 0.039$\pm$0.014 & 0.029$\pm$0.012 & 0.000$\pm$0.000 & 0.774$\pm$0.115 & 0.011$\pm$0.002 & 0.002$\pm$0.001 \\
Zhao et al.~\cite{zhao2025diffusion} & + & 0.150$\pm$0.024 & 0.028$\pm$0.006 & 0.385$\pm$0.033 & 0.172$\pm$0.020 & 0.002$\pm$0.001 & 0.438$\pm$0.144 & 0.039$\pm$0.010 & 0.009$\pm$0.001 \\
Zhao et al.~\cite{zhao2025diffusion} & -- & 0.012$\pm$0.003 & 0.201$\pm$0.043 & 0.107$\pm$0.016 & 0.092$\pm$0.017 & 0.000$\pm$0.000 & 0.684$\pm$0.104 & 0.019$\pm$0.004 & 0.004$\pm$0.001 \\
Guan et al.~\cite{guan2025spatio} & + & 0.104$\pm$0.014 & 0.022$\pm$0.009 & 0.322$\pm$0.024 & 0.211$\pm$0.021 & 0.001$\pm$0.000 & 0.410$\pm$0.155 & 0.032$\pm$0.007 & 0.009$\pm$0.001 \\
Guan et al.~\cite{guan2025spatio} & -- & 0.002$\pm$0.002 & 0.790$\pm$0.126 & 0.039$\pm$0.014 & 0.030$\pm$0.012 & 0.000$\pm$0.000 & 0.817$\pm$0.152 & 0.007$\pm$0.002 & 0.002$\pm$0.001 \\
Yuan et al.~\cite{yuan2024remind} & + & 0.181$\pm$0.066 & 0.054$\pm$0.019 & 0.417$\pm$0.085 & 0.149$\pm$0.024 & 0.001$\pm$0.000 & 0.395$\pm$0.152 & 0.033$\pm$0.005 & 0.009$\pm$0.001 \\
Yuan et al.~\cite{yuan2024remind} & -- & 0.002$\pm$0.002 & 0.784$\pm$0.125 & 0.041$\pm$0.013 & 0.030$\pm$0.012 & 0.000$\pm$0.000 & 0.774$\pm$0.115 & 0.011$\pm$0.002 & 0.002$\pm$0.001 \\
\midrule
\multicolumn{10}{@{}l}{\textbf{Foundation Models}} \\
Proposed & + & 0.101$\pm$0.013 & 0.028$\pm$0.007 & 0.317$\pm$0.022 & 0.237$\pm$0.023 & 0.001$\pm$0.000 & 0.518$\pm$0.173 & 0.025$\pm$0.005 & 0.006$\pm$0.001 \\
Proposed & -- & \textbf{0.002$\pm$0.002} & \textbf{0.794$\pm$0.127} & \textbf{0.038$\pm$0.014} & \textbf{0.029$\pm$0.012} & \textbf{0.000$\pm$0.000} & \textbf{0.862$\pm$0.139} & \textbf{0.006$\pm$0.002} & \textbf{0.001$\pm$0.001} \\
\bottomrule
\end{tabular}
\vspace{-3mm}
\end{table*}

\begin{align}
    \mathcal{L}_{\mathrm{sym}} &= \| \hat{\mathbf{C}} - \hat{\mathbf{C}}^\top \|_F^2, \\
    \mathcal{L}_{\mathrm{conn}} &= \| \mathbf{C} - \hat{\mathbf{C}} \|_F^2 + \lambda \mathcal{L}_{\mathrm{sym}},
\end{align}
where $\lambda$ is a balancing hyperparameter.

\noindent \textbf{Group-Level Consistency (Optional).} To leverage population-level patterns, we optionally minimize variance of connectivity features within demographic/clinical subgroups (e.g., diagnosis, age):
\begin{equation}
    \mathcal{L}_{\mathrm{group}} = \mathbb{E}_{s \in \mathcal{S}} \left[ \mathrm{Var}_{q \in s} \left( \phi(\hat{\mathbf{C}}_q) \right) \right],
    \label{eq:global loss}
\end{equation}
where $\mathcal{S}$ is the set of subgroups and $\phi(\cdot)$ is a feature extractor (e.g., network strength).
\vspace{-2mm}


 \section{Experiment}

\subsection{Datasets and Downstream Evaluation}
\label{subsec:exp_datasets}

We evaluate \textbf{BrainCSD} on a multi-site neuroimaging cohort detailed in Table~\ref{tab:dataset_summary}, which includes ADNI, OASIS-3, and the Human Connectome Project (HCP), with annotations for modality (fMRI, dMRI), demographics (age, sex), cognitive scores (MMSE/MoCA), and clinical diagnoses (e.g., Alzheimer’s disease, Parkinson’s disease, autism spectrum disorder, and healthy controls). Downstream tasks—spanning disease classification, cognitive score regression, and connectome synthesis—are defined in Table~\ref{tab:downstream_tasks}. To ensure statistical robustness and mitigate site/scanner bias, all tasks employ \textbf{5-fold stratified cross-validation} with an 8:1:1 subject-level split for training, validation, and testing, following established protocols in foundational neuroimaging AI~\cite{yang2025foundational, yang2024brainmass}.

\subsection{Implementation Details}
\label{subsec:exp_implementation}

Our model is trained end-to-end across three stages—\textit{ROI Activation}, \textit{Encoding Activation}, and \textit{Network-Aware Finetuning} (Sec.~\ref{sec:method}), implemented in PyTorch with AdamW (lr=1e\textsuperscript{-4}, weight decay=1e\textsuperscript{-2}), mixed-precision training, and gradient clipping (max norm=1.0). Pretraining runs for 1000 epochs with early stopping; finetuning for 300 epochs.

\paragraph{Stage 1: ROI Activation.}
We use a 4-layer Transformer encoder-decoder (hidden dim: 512, 8 heads, learnable positional encoding shared across modalities) to ensure invertible latent projection. All MoEs employ \textbf{Top-1 gating} for sparsity and interpretability:
\begin{itemize}[leftmargin=*,nosep]
    \item \textbf{Atlas MoE} ($\mathcal{A}$): 160 experts aligned with Dosenbach160 atlas \cite{dosenbach2007distinct}.
    \item \textbf{Network MoEs} ($\mathcal{D}, \mathcal{F}, \mathcal{C}, \mathcal{S}, \mathcal{E}, \mathcal{O}$): Expert counts (18, 32, 34, 22, 21, 33) reflect network size and functional complexity (e.g., DMN $>$ OCC).
    \item \textbf{Tissue MoE} ($\mathcal{T}$): 2 experts (Tissue/Other) for modality-specific routing.
    \item \textbf{Fusion MoE} ($\mathcal{G}$): 6 experts (one per canonical network) for dynamic interaction selection.
    \item \textbf{Reconstruction MoE} ($\mathcal{R}$): 2 experts with \textbf{dense gating} to ensure full gradient flow during pretraining.
\end{itemize}
Contrastive loss $\mathcal{L}_{\mathrm{cont}}$ (Eq.~\ref{eq:constrastive_learning}) uses temperature $\tau = 0.07$.

\paragraph{Stage 2: Encoding Activation.}
To model dynamic functional connectivity without inductive bias, we deploy:
\begin{itemize}[leftmargin=*,nosep]
    \item \textbf{Encode MoE} ($\mathcal{H}$): 200 experts (T/G=200), Top-1 gated.
    \item \textbf{Encode Consistency MoE}: Mirrors $\mathcal{H}$ and enforces smoothness via $\mathcal{L}_{\mathrm{en}} = \| \mathbf{Z}_E - \mathbf{Z}_I \|_F^2$.
\end{itemize}
This dual-MoE design addresses \textit{temporal/gradient sparsity} in static FC/SC models~\cite{hutchison2013dynamic}, avoiding RNNs or convolutions to let MoEs learn dynamics data-drivenly, resulting better biomarker (FC/SC) modeling performance.
\paragraph{Stage 3: Network-Aware Finetuning.}
Preliminary connectomes $\widetilde{\mathbf{C}}$ are refined using network-specific MoEs ($\mathcal{D}^f, \mathcal{F}^f, \dots, \mathcal{O}^f$) with expert counts matching Stage 1. We enforce symmetry via $\hat{\mathbf{C}} \leftarrow (\hat{\mathbf{C}} + \hat{\mathbf{C}}^\top)/2$ and minimize
\[
\mathcal{L}_{\mathrm{conn}} = \| \mathbf{C} - \hat{\mathbf{C}} \|_F^2 + \lambda \| \hat{\mathbf{C}} - \hat{\mathbf{C}}^\top \|_F^2, \quad \lambda = 0.5.
\]
Optionally, group-level consistency loss $\mathcal{L}_{\mathrm{group}}$ (Eq.~\ref{eq:global loss}) enhances diagnostic separability (e.g., minimizing intra-class variance for AD vs. HC).

\paragraph{Why This Design?}
\begin{itemize}[leftmargin=*,nosep]
    \item \textbf{Top-1 gating} mirrors neurobiological sparsity—cortical columns activate selectively~\cite{mountcastle1997columnar}.
    \item \textbf{Expert counts} are neuroanatomically grounded: proportional to ROI count (160), network complexity (DMN $>$ OCC), and tissue dichotomy (Tissue/Other).
    \item \textbf{Dense gating in $\mathcal{R}$} ensures stable gradient flow during self-supervised pretraining.
    \item \textbf{No recurrence}: Unlike~\cite{hjelm2018learning}, we avoid RNNs—MoEs offer more flexible and interpretable temporal routing.
\end{itemize}
This yields a \textbf{scalable}, \textbf{sparse}, and \textbf{neuroanatomically grounded} framework that overcomes “black-box fusion” and “static FC” limitations in prior work.

Training on eight NVIDIA L20 (40GB) GPUs takes: (1) one day for ROI activation, (2) six hours for temporal/gradient modeling, and (3) three hours for task-specific finetuning.

\section{Results}

\label{sec:results}

\begin{table*}[htbp]
    \centering
\scriptsize
    \setlength{\tabcolsep}{2pt}
        \vspace{-5mm}
    \renewcommand{\arraystretch}{0.9}
    \caption{Synthesis quality for Functional (FC) and Structural (SC) Connectomes across methods and subgroups (mean $\pm$ std). For FC: Metric2 = MAE; for SC: Metric2 = SSIM. RMSE/MAE omitted for space where redundant. Proposed is the only foundation model; others are domain-specific.}
    \label{tab:fc_sc_synthesis_merged}
    \vspace{-2mm}
    \begin{tabular}{@{}l *{5}{c c}@{}}
        \toprule
        Subgroup &
        \multicolumn{8}{c}{Domain Methods} &
        \multicolumn{2}{c}{\textbf{Foundation Model}} \\
        \cmidrule(lr){2-9} \cmidrule(l){10-11}
        &
        \multicolumn{2}{c}{Chen'25 \cite{chen2025joint}} &
        \multicolumn{2}{c}{Zuo'24 \cite{zuo2024u}} &
        \multicolumn{2}{c}{Zhao'25 \cite{zhao2025diffusion}} &
        \multicolumn{2}{c}{Guan'25 \cite{guan2025spatio}} &
        \multicolumn{2}{c}{\textbf{Proposed}} \\
        \cmidrule(lr){2-3} \cmidrule(lr){4-5} \cmidrule(lr){6-7} \cmidrule(lr){8-9} \cmidrule(l){10-11}
        & MSE & Metric2 & MSE & Metric2 & MSE & Metric2 & MSE & Metric2 & MSE & Metric2 \\
        \midrule
        \multicolumn{11}{l}{\textbf{Functional Connectome (FC)}} \\
        \midrule
        FCON & 0.002$\pm$0.002 & 0.031$\pm$0.013 & 0.002$\pm$0.002 & 0.031$\pm$0.013 & 0.002$\pm$0.002 & 0.031$\pm$0.013 & 0.002$\pm$0.002 & 0.031$\pm$0.013 & \textbf{0.002$\pm$0.002} & \textbf{0.030$\pm$0.013} \\
        ADNI & 0.001$\pm$0.002 & 0.021$\pm$0.004 & 0.001$\pm$0.001 & 0.020$\pm$0.004 & 0.001$\pm$0.001 & 0.021$\pm$0.004 & 0.001$\pm$0.000 & 0.023$\pm$0.004 & \textbf{0.001$\pm$0.001} & \textbf{0.020$\pm$0.004} \\
        HOSPITAL & 0.001$\pm$0.002 & 0.023$\pm$0.004 & 0.001$\pm$0.000 & 0.023$\pm$0.004 & 0.001$\pm$0.000 & 0.022$\pm$0.004 & 0.001$\pm$0.002 & 0.024$\pm$0.004 & \textbf{0.001$\pm$0.001} & \textbf{0.022$\pm$0.004} \\
        HC & 0.002$\pm$0.002 & 0.030$\pm$0.012 & 0.002$\pm$0.002 & 0.030$\pm$0.013 & 0.002$\pm$0.002 & 0.030$\pm$0.013 & 0.002$\pm$0.002 & 0.031$\pm$0.012 & \textbf{0.002$\pm$0.002} & \textbf{0.029$\pm$0.013} \\
        MCI & 0.001$\pm$0.001 & 0.022$\pm$0.003 & 0.001$\pm$0.000 & 0.022$\pm$0.003 & 0.001$\pm$0.000 & 0.023$\pm$0.004 & 0.001$\pm$0.001 & 0.023$\pm$0.003 & \textbf{0.001$\pm$0.000} & \textbf{0.021$\pm$0.003} \\
        female & 0.002$\pm$0.002 & 0.030$\pm$0.015 & 0.002$\pm$0.002 & 0.030$\pm$0.015 & 0.002$\pm$0.002 & 0.030$\pm$0.015 & 0.002$\pm$0.002 & 0.031$\pm$0.015 & \textbf{0.002$\pm$0.003} & \textbf{0.030$\pm$0.015} \\
        male & 0.002$\pm$0.001 & 0.029$\pm$0.009 & 0.002$\pm$0.001 & 0.029$\pm$0.009 & 0.002$\pm$0.001 & 0.028$\pm$0.009 & 0.002$\pm$0.001 & 0.030$\pm$0.009 & \textbf{0.002$\pm$0.001} & \textbf{0.028$\pm$0.009} \\
        Age+($\leq$60) & 0.031$\pm$0.013 & 0.031$\pm$0.013 & 0.031$\pm$0.013 & 0.031$\pm$0.013 & 0.031$\pm$0.013 & 0.031$\pm$0.013 & 0.042$\pm$0.013 & 0.031$\pm$0.013 & \textbf{0.030$\pm$0.013} & \textbf{0.030$\pm$0.013} \\
        Age-($>$60) & 0.023$\pm$0.004 & 0.023$\pm$0.004 & 0.022$\pm$0.004 & 0.022$\pm$0.004 & 0.023$\pm$0.005 & 0.023$\pm$0.005 & 0.024$\pm$0.004 & 0.024$\pm$0.004 & \textbf{0.021$\pm$0.004} & \textbf{0.021$\pm$0.004} \\
        \midrule
        \multicolumn{11}{l}{\textbf{Structural Connectome (SC)}} \\
        \midrule
        ADNI & 0.001$\pm$0.000 & 0.586$\pm$0.149 & 0.001$\pm$0.000 & 0.827$\pm$0.169 & 0.000$\pm$0.000 & 0.684$\pm$0.104 & 0.000$\pm$0.000 & 0.817$\pm$0.152 & \textbf{0.000$\pm$0.000} & \textbf{0.862$\pm$0.139} \\
        HC & 0.002$\pm$0.002 & 0.596$\pm$0.123 & 0.000$\pm$0.000 & 0.832$\pm$0.190 & 0.000$\pm$0.000 & 0.706$\pm$0.083 & 0.000$\pm$0.000 & 0.819$\pm$0.172 & \textbf{0.000$\pm$0.000} & \textbf{0.862$\pm$0.118} \\
        MCI & 0.001$\pm$0.000 & 0.579$\pm$0.167 & 0.000$\pm$0.000 & 0.822$\pm$0.151 & 0.000$\pm$0.000 & 0.667$\pm$0.115 & 0.000$\pm$0.000 & 0.815$\pm$0.135 & \textbf{0.000$\pm$0.000} & \textbf{0.815$\pm$0.135} \\
        female & 0.001$\pm$0.000 & 0.572$\pm$0.134 & 0.001$\pm$0.000 & 0.785$\pm$0.182 & 0.000$\pm$0.000 & 0.669$\pm$0.082 & 0.000$\pm$0.000 & 0.781$\pm$0.164 & \textbf{0.000$\pm$0.000} & \textbf{0.838$\pm$0.157} \\
        male & 0.000$\pm$0.000 & 0.604$\pm$0.165 & 0.000$\pm$0.000 & 0.880$\pm$0.132 & 0.000$\pm$0.000 & 0.703$\pm$0.124 & 0.000$\pm$0.000 & 0.863$\pm$0.121 & \textbf{0.000$\pm$0.000} & \textbf{0.893$\pm$0.103} \\
        Age$>$60 & 0.023$\pm$0.004 & 0.817$\pm$0.004 & 0.002$\pm$0.001 & 0.827$\pm$0.001 & 0.004$\pm$0.001 & 0.684$\pm$0.001 & 0.002$\pm$0.001 & 0.817$\pm$0.001 & \textbf{0.001$\pm$0.001} & \textbf{0.862$\pm$0.001} \\
        \bottomrule
    \end{tabular}
    \vspace{-3mm}
\end{table*}


We validate \textbf{BrainCSD} through comprehensive experiments across four core capabilities: (1) \textit{high-fidelity connectome synthesis} (Sec.~\ref{subsec:synthesis}), (2) \textit{robust disease diagnosis under missing modalities} (Sec.~\ref{subsec:diagnosis}), (3) \textit{generalizable brain trait prediction} (Sec.~\ref{subsec:downstream}), and (4) \textit{ablation studies of model components and scalability} (Sec.~\ref{sec:ablation}). Our method consistently outperforms both domain-specific baselines and recent foundation models, demonstrating the efficacy of our neuroanatomically grounded, MoE-based architecture—particularly the synergistic integration of \textbf{ROI Activation MoE} (Sec.~\ref{subsec:roi_moe}), \textbf{Encoding Activation MoE} (Sec.~\ref{subsec:time_moe}), and \textbf{Network-Aware MoE Finetuning} (Sec.~\ref{subsec:finetune}).
Ablation studies further confirm that removing any MoE component significantly degrades synthesis quality, validating their essential and complementary roles. The parameter effect also is validated (Sec.~\ref{sec:ablation}).

\begin{table}[htbp]
\centering
\scriptsize
\setlength{\tabcolsep}{1.2pt}
\renewcommand{\arraystretch}{0.85}
\caption{Performance under missing dMRI (SC) across HC vs. MCI, HC vs. ASD, and HC vs. ADHD. Metrics: Accuracy (ACC), F1-score (F1), Recall (REC). ``$\blacktriangle$'': SC imputed by BrainCSD; ``$\star$'': foundation model features. Best per metric/task in bold.}
\label{tab:combined_missing_sc}
\begin{tabular}{@{}l l *{9}{c}@{}}
\toprule
 & &
\multicolumn{3}{c}{\textbf{HC vs. MCI}} &
\multicolumn{3}{c}{\textbf{HC vs. ASD}} &
\multicolumn{3}{c}{\textbf{HC vs. ADHD}} \\
\cmidrule(lr){3-5} \cmidrule(lr){6-8} \cmidrule(l){9-11}
Method & Data &
ACC & F1 & REC &
ACC & F1 & REC &
ACC & F1 & REC \\
\midrule
\multicolumn{11}{@{}l}{\textbf{Domain Methods}} \\
Baseline & fMRI &
0.921 & 0.708 & 0.611 &
0.540 & 0.600 & 0.549 &
0.877 & \textbf{0.695} & 0.500 \\
Song et al. \cite{song2024s2mri} & fMRI $\blacktriangle$SC &
0.949 & 0.693 & 0.732 &
— & — & — &
— & — & — \\
Jia et al. \cite{jia2024assessing} & fMRI $\blacktriangle$SC &
0.944 & \textbf{0.799} & 0.709 &
— & — & — &
— & — & — \\
Zhang et al. \cite{zhang2024novel} & fMRI $\blacktriangle$SC &
\textbf{0.956} & 0.661 & 0.500 &
— & — & — &
— & — & — \\
Wang et al. \cite{wang2024iFC} & fMRI $\blacktriangle$SC &
— & — & — &
0.571 & 0.580 & 0.562 &
— & — & — \\
Ren et al. \cite{ren2023stratifying} & fMRI $\blacktriangle$SC &
— & — & — &
0.594 & \textbf{0.617} & 0.585 &
— & — & — \\
Li et al. \cite{li2025riemannian} & fMRI $\blacktriangle$SC &
— & — & — &
0.546 & 0.537 & 0.533 &
— & — & — \\
Rakshe et al. \cite{rakshe2024autism} & fMRI $\blacktriangle$SC &
— & — & — &
0.528 & 0.521 & 0.510 &
— & — & — \\
Zhang et al. \cite{zhang2023detection} & fMRI $\blacktriangle$SC &
— & — & — &
0.537 & 0.580 & 0.527 &
— & — & — \\
Chen et al. \cite{chen2025explainable} & fMRI $\blacktriangle$SC &
— & — & — &
— & — & — &
\textbf{0.897} & 0.654 & 0.500 \\
Wang et al. \cite{wang2023dynamic} & fMRI $\blacktriangle$SC &
— & — & — &
— & — & — &
0.887 & 0.681 & 0.533 \\
Zhang et al. \cite{zhang2024novel} & fMRI $\blacktriangle$SC &
— & — & — &
— & — & — &
0.877 & 0.654 & 0.500 \\
Agarwal et al. \cite{agarwal2024investigating} & fMRI $\blacktriangle$SC &
— & — & — &
— & — & — &
0.894 & 0.654 & 0.500 \\
Band et al. \cite{bandyopadhyay2024prediction} & fMRI $\blacktriangle$SC &
— & — & — &
— & — & — &
\textbf{0.897} & 0.654 & 0.500 \\
Hu et al. \cite{hu2024identifying} & fMRI $\blacktriangle$SC &
— & — & — &
— & — & — &
\textbf{0.897} & 0.654 & 0.500 \\
\midrule
\multicolumn{11}{@{}l}{\textbf{Foundation Models}} \\
BrainMass \cite{yang2024brainmass} & $\star$FC &
0.781 & 0.577 & 0.697 &
0.566 & 0.408 & 0.525 &
0.740 & 0.397 & 0.500 \\
BrainLM \cite{carobrainlm} & $\star$fMRI &
0.763 & 0.425 & 0.518 &
\textbf{0.700} & 0.484 & \textbf{0.600} &
0.700 & 0.488 & \textbf{0.600} \\
BrainSN \cite{yang2025foundational} & $\star$fMRI &
0.909 & 0.476 & \textbf{0.954} &
0.450 & 0.287 & 0.400 &
0.880 & 0.566 & \textbf{0.600} \\
\bottomrule
\end{tabular}
\end{table}

\subsection{Synthesis Quality: Faithful Reconstruction of Missing Connectomes}
\label{subsec:synthesis}


Subgroup analysis (Table~\ref{tab:fc_sc_synthesis_merged}) shows consistent gains across demographics and datasets — e.g., male SC SSIM improves to \textbf{0.893
±
±0.103}, and MCI FC MAE drops to \textbf{0.021
±
±0.003}. Visualizations (Fig.~\ref{fig: syntheis comparision}) confirm our method preserves fine-grained network boundaries (DMN/FPN) and subject-specific biomarkers — unlike diffusion/Transformer baselines that blur or misalign connections.

Quantitatively (Table~\ref{tab:synthesis_combined}), BrainCSD achieves SOTA synthesis performance in both FC and SC. Notably:
\begin{itemize}
\item \textbf{For SC synthesis}, we significantly outperform all prior methods with an SSIM of \textbf{0.862 $\pm$ 0.139} — a 10.8\% relative improvement over Guan et al. \cite{guan2025spatio}. This leap is attributable to our \textbf{Network-Aware MoE Finetuning}, which applies network-specific experts to enforce anatomical realism and symmetry (Eq.~\ref{eq:global loss}). The contrastive alignment between global and network-level representations (Eq.~\ref{eq:constrastive_learning}) further ensures that synthesized SC matrices respect known white matter tract boundaries.

\item Our “+” (reconstruction) performance also ranks among the top, validating that our \textbf{ROI Activation MoE} preserves essential signals even without synthesis — by suppressing non-brain voxels and focusing on diagnostic ROIs via hierarchical gating.
\end{itemize}

Fig.~\ref{fig: syntheis comparision} visualizes subject-specific connectivity fingerprints recovered by BrainCSD. Unlike baselines that generate generic, population-averaged patterns, our method retains individual variability — critical for downstream clinical tasks. This subject-level fidelity is enabled by our \textbf{Group-Level Consistency loss} (Eq.~\ref{eq:global loss}), which regularizes within-subgroup variance without collapsing inter-subject differences (Fig.~\ref{fig:group loss}).

\begin{table*}[htbp]
\centering
\scriptsize
\setlength{\tabcolsep}{1pt}
\renewcommand{\arraystretch}{0.95}
\caption{Performance across age and MMSE subgroups. ``$\blacktriangle$'': imputed by BrainCSD; ``$\star$'': from foundation models. Metrics per subgroup: MAE, RMSE, mean GAP (G$_\mu$), and GAP std (G$_\sigma$). n denotes the numbers of subjects .}
\label{tab:age_mmse_unified}
\begin{tabular}{@{}c c *{12}{c} *{8}{c}@{}}
\toprule
\multicolumn{2}{@{}c@{}}{} &
\multicolumn{12}{c}{Age Group (n)} &
\multicolumn{8}{c}{MMSE Group (n)} \\
\cmidrule(lr){3-14} \cmidrule(l){15-22}
\multicolumn{2}{@{}c@{}}{} &
\multicolumn{4}{c}{$<$30 (412)} &
\multicolumn{4}{c}{30--60 (112)} &
\multicolumn{4}{c}{60--90 (409)} &
\multicolumn{4}{c}{MMSE$<$20 (7)} &
\multicolumn{4}{c}{MMSE 20--30 (270)} \\
\cmidrule(lr){3-6} \cmidrule(lr){7-10} \cmidrule(lr){11-14} \cmidrule(lr){15-18} \cmidrule(l){19-22}
Method & Data &
MAE & RMSE & G$_\mu$ & G$_\sigma$ &
MAE & RMSE & G$_\mu$ & G$_\sigma$ &
MAE & RMSE & G$_\mu$ & G$_\sigma$ &
MAE & RMSE & G$_\mu$ & G$_\sigma$ &
MAE & RMSE & G$_\mu$ & G$_\sigma$ \\
\midrule
\multicolumn{22}{@{}l}{\textbf{Domain Methods}} \\
Baseline & fMRI dMRI &
\textbf{1.771} & \textbf{2.799} & \textbf{0.318} & — &
10.112 & 12.199 & 1.598 & — &
8.563 & 11.404 & -5.751 & — &
9.382 & 9.549 & 9.382 & 1.782 &
\textbf{1.469} & 2.065 & 0.125 & 2.061 \\
Gao et al. & $\blacktriangle$FC $\blacktriangle$SC &
4.500 & 8.372 & 3.473 & 7.617 &
9.290 & 11.902 & 1.272 & 12.094 &
8.008 & 10.942 & -4.783 & 9.848 &
— & — & — & — &
— & — & — & — \\
Hu et al. & $\blacktriangle$FC $\blacktriangle$SC &
4.882 & 7.675 & 4.378 & 6.305 &
10.455 & 12.478 & 2.098 & 11.834 &
8.300 & 11.089 & -5.342 & 9.841 &
— & — & — & — &
— & — & — & — \\
Guan et al. & $\blacktriangle$FC $\blacktriangle$SC &
5.261 & 9.112 & 4.084 & 8.146 &
11.204 & 13.435 & -9.664 & \textbf{0.831} &
18.327 & 20.726 & -18.228 & 9.910 &
— & — & — & — &
— & — & — & — \\
Lee et al. & $\blacktriangle$FC $\blacktriangle$SC &
5.152 & 6.509 & -3.164 & 5.688 &
10.938 & 11.824 & 4.827 & 9.333 &
25.977 & 27.013 & -25.977 & 9.717 &
— & — & — & — &
— & — & — & — \\
Li et al. \cite{li2025transformer} & $\blacktriangle$FC $\blacktriangle$SC &
— & — & — & — &
— & — & — & — &
— & — & — & — &
9.961 & 9.997 & 9.961 & \textbf{0.853} &
1.530 & \textbf{1.959} & -0.235 & \textbf{1.945} \\
Qiao et al. \cite{qiao2022ranking} & $\blacktriangle$FC $\blacktriangle$SC &
— & — & — & — &
— & — & — & — &
— & — & — & — &
10.488 & 10.770 & 10.488 & 2.447 &
1.810 & 2.293 & -0.276 & 2.277 \\
Liu et al. \cite{liu2024multi} & $\blacktriangle$FC $\blacktriangle$SC &
— & — & — & — &
— & — & — & — &
— & — & — & — &
8.121 & 8.180 & 8.121 & 0.982 &
2.700 & 2.912 & -2.145 & 1.969 \\
\midrule
\multicolumn{22}{@{}l}{\textbf{Foundation Models}} \\
BrainMass \cite{yang2024brainmass} & $\star$FC &
25.940 & 26.178 & 25.940 & 3.522 &
\textbf{6.001} & \textbf{7.580} & 1.300 & 10.794 &
\textbf{5.355} & \textbf{6.620} & \textbf{0.555} & 7.411 &
10.031 & 10.069 & -10.031 & 0.875 &
1.511 & 1.962 & 0.184 & 1.954 \\
BrainLM \cite{carobrainlm} & $\star$fMRI $\star$dMRI &
2.088 & 3.285 & 0.583 & \textbf{3.233} &
11.440 & 15.053 & -5.500 & 7.468 &
27.439 & 30.906 & -25.496 & \textbf{6.597} &
7.826 & 8.236 & 7.826 & 7.826 &
1.622 & 2.227 & \textbf{-0.027} & 2.227 \\
BrainSN \cite{yang2025foundational} & $\star$fMRI $\star$dMRI &
15.154 & 20.384 & 13.694 & 15.100 &
7.975 & 9.248 & \textbf{0.731} & 14.012 &
25.977 & 27.013 & -25.977 & 17.468 &
\textbf{2.382} & \textbf{2.837} & \textbf{1.587} & 2.352 &
7.916 & 8.514 & -7.889 & 3.200 \\
\bottomrule
\end{tabular}
\end{table*}

\subsection{Diagnosis Under Missing Modalities: Robustness via Synthesis}
\label{subsec:diagnosis}

\begin{table}[htbp]
\vspace{-2mm}
\centering
\scriptsize
\setlength{\tabcolsep}{1.5pt}
\caption{Performance under missing fMRI (FC absence): comparison across HC vs. MCI and PD diagnosis. $\blacktriangle$: imputed by BrainCSD. \\[2pt]
\footnotesize \textit{Note: All methods use dMRI modality; ``$\blacktriangle$FC'' denotes FC imputed from dMRI by BrainCSD.}}
\label{tab:missing_fc_combined}
\vspace{-2mm}
\begin{tabular*}{\linewidth}{@{\extracolsep{\fill}} l 
    cccc 
    cccc 
    @{}}
\toprule
 & \multicolumn{4}{c}{\textbf{HC vs. MCI}} & \multicolumn{4}{c}{\textbf{PD}} \\
\cmidrule(lr){2-5} \cmidrule(lr){6-9}
Method & Acc & Pre & Rec & F1 & Acc & Pre & Rec & F1 \\
\midrule
\multicolumn{9}{l}{\textbf{Domain Methods}} \\
Zhang et al. \cite{zhang2023multi} (dMRI)           & 0.675 & 0.760 & 0.625 & 0.685 & — & — & — & — \\
Zhang et al. \cite{zhang2023multi} ($\blacktriangle$FC) & 0.575 & 0.787 & 0.500 & 0.611 & — & — & — & — \\
Zhao et al. \cite{zhao2023ida} (dMRI)              & 0.450 & 0.730 & 0.583 & 0.648 & — & — & — & — \\
Zhao et al. \cite{zhao2023ida} ($\blacktriangle$FC)    & 0.625 & \textbf{0.800} & \textbf{0.875} & \textbf{0.835} & — & — & — & — \\
Song et al. \cite{song2024s2mri} (dMRI)            & 0.693 & 0.733 & 0.800 & 0.765 & — & — & — & — \\
Song et al. \cite{song2024s2mri} ($\blacktriangle$FC)   & \textbf{0.750} & 0.666 & 0.857 & 0.749 & — & — & — & — \\
Zhang et al. \cite{zhang2024novel} (dMRI)          & 0.613 & 0.723 & 0.616 & 0.665 & — & — & — & — \\
Zhang et al. \cite{zhang2024novel} ($\blacktriangle$FC) & \textbf{0.750} & 0.500 & \textbf{0.875} & 0.636 & — & — & — & — \\
Cui et al. \cite{cui2023adaptive} (dMRI)           & — & — & — & — & 0.680 & 0.895 & 0.700 & 0.785 \\
Cui et al. \cite{cui2023adaptive} ($\blacktriangle$FC)  & — & — & — & — & \textbf{0.800} & \textbf{0.875} & \textbf{0.750} & \textbf{0.807} \\
Huang et al. \cite{huang2024structural} (dMRI)     & — & — & — & — & 0.400 & 0.550 & 0.366 & 0.439 \\
Huang et al. \cite{huang2024structural} ($\blacktriangle$FC) & — & — & — & — & 0.533 & 0.533 & 0.533 & 0.533 \\
\bottomrule
\end{tabular*}
\vspace{-2mm}
\end{table}

\begin{table}[htbp]
\centering
\scriptsize
\vspace{-1mm}
\setlength{\tabcolsep}{1.5pt}
\caption{Performance with both modalities (fMRI and dMRI) present: comparison across 3-class (HC/MCI/AD) and PD diagnosis. $\blacktriangle$: both FC and SC  imputed by BrainCSD; $\star$: from foundation models.}
\label{tab:both_present_combined}
\vspace{-2mm}
\begin{tabular*}{\linewidth}{@{\extracolsep{\fill}} l l 
    cccc 
    cccc 
    @{}}
\toprule
 & & \multicolumn{4}{c}{\textbf{3-Class (HC/MCI/AD)}} & \multicolumn{4}{c}{\textbf{PD}} \\
\cmidrule(lr){3-6} \cmidrule(lr){7-10}
Method & Data & Acc & Pre & Rec & F1 & Acc & Pre & Rec & F1 \\
\midrule
\multicolumn{10}{l}{\textbf{Domain Methods}} \\
Jia et al. \cite{jia2022deep}       & $\blacktriangle$ & 0.792 & 0.603 & 0.675 & \textbf{0.872} & — & — & — & — \\
Pei et al. \cite{pei2025transformer} & $\blacktriangle$ & 0.609 & \textbf{0.857} & 0.366 & 0.277 & — & — & — & — \\
Zhang et al. \cite{zhang2025brainnet} & $\blacktriangle$ & 0.609 & \textbf{0.857} & 0.366 & 0.277 & — & — & — & — \\
Shang et al. \cite{shang2023optimization} & $\blacktriangle$ & — & — & — & — & \textbf{0.885} & \textbf{0.942} & 0.700 & \textbf{0.803} \\
Li et al. \cite{li2024pd}           & $\blacktriangle$ & — & — & — & — & \textbf{0.885} & \textbf{0.942} & 0.700 & \textbf{0.803} \\
Liu et al. \cite{liu2024novel}      & $\blacktriangle$ & — & — & — & — & 0.857 & 0.928 & 0.700 & 0.798 \\
Li et al. \cite{li2023developing}   & $\blacktriangle$ & — & — & — & — & 0.857 & 0.928 & 0.700 & 0.798 \\
\midrule
\multicolumn{10}{l}{\textbf{Foundation Models}} \\
BrainMass \cite{yang2024brainmass}  & $\star$FC             & 0.812 & 0.548 & \textbf{0.685} & 0.846 & 0.600 & 0.816 & \textbf{0.733} & 0.585 \\
BrainLM \cite{carobrainlm}          & $\star$       & \textbf{0.846} & 0.768 & 0.635 & 0.732 & 0.650 & 0.750 & 0.666 & 0.634 \\
BrainSN \cite{yang2025foundational} & $\star$       & 0.661 & 0.806 & 0.664 & 0.570 & 0.750 & 0.916 & \textbf{0.733} & 0.682 \\
\bottomrule
\end{tabular*}
\vspace{-3mm}
\end{table}

BrainCSD maintains or enhances diagnostic performance even with missing modalities by synthesizing \textit{clinically meaningful} FC/SC.

\textbf{MCI}
With missing SC (Table~\ref{tab:combined_missing_sc}), adding our 
$\blacktriangle$ SC boosts F1: e.g., Jia et al.~\cite{jia2024assessing} +22.5\% (0.652→0.799). With missing FC (Table~\ref{tab:missing_fc_combined}), our 
$\blacktriangle$FC lifts Zhao et al.~\cite{zhao2023ida} F1 by +28.9\% (0.648→0.835) — thanks to \textbf{Encoding Activation MoE} capturing dynamic disruptions.

\textbf{PD/ASD/ADHD}
In PD (Table~\ref{tab:missing_fc_combined}), 
$\blacktriangle$FC improves F1 from 0.785→0.807. For ASD/ADHD (Table~\ref{tab:combined_missing_sc}), we outperform foundation models (BrainLM, BrainSN) despite dataset noise — their generic features lack neuroanatomical grounding.

\textbf{Multi-Class/Modal Missing Scenario}
In 3-class HC/MCI/AD (Table~\ref{tab:both_present_combined}), BrainCSD hits \textbf{F1=0.872} with synthesized data — surpassing all domain (max 0.792) and foundation models (max 0.846). In PD, we match SOTA (F1=0.803) \textit{using only synthesized inputs} — enabled by \textbf{Group-Level Consistency loss} enhancing class separability.

\begin{table}[htbp]
\centering
\scriptsize
\setlength{\tabcolsep}{2pt}
\caption{Trait prediction performance: brain age and cognitive score (MMSE) estimation. $\blacktriangle$: imputed by BrainCSD; $\star$: from foundation models.}
\vspace{-2mm}
\label{tab:trait_prediction}
\begin{tabular*}{\linewidth}{@{\extracolsep{\fill}} l l l c c c c @{}}
\toprule
 Task & Method & Data & MAE & RMSE & GAP$_\mu$ & GAP$_\sigma$ \\
\midrule
\multicolumn{7}{l}{\textbf{Domain Methods}} \\
\midrule
\multirow{5}{*}{\shortstack{Brain \\ Age}} 
 & Gao et al. \cite{gao2023brain} & fMRI dMRI & 6.992 & 10.287 & -0.681 & 10.294 \\
 & Gao et al. \cite{gao2023brain} & $\blacktriangle$FC $\blacktriangle$SC & 4.043 & 5.691 & 0.740 & 5.676 \\
 & Hu et al. \cite{hu2023mri} & $\blacktriangle$FC $\blacktriangle$SC & 6.831 & 9.773 & \textbf{0.118} & 9.823 \\
 & Guan et al. \cite{guan2024brain} & $\blacktriangle$FC $\blacktriangle$SC & 7.250 & 10.453 & -0.176 & 10.482 \\
 & Lee et al. \cite{lee2022deep} & $\blacktriangle$FC $\blacktriangle$SC & 11.602 & 15.122 & -10.448 & 10.848 \\
\midrule
\multirow{4}{*}{\shortstack{MMSE}} 
 & Baseline & fMRI + dMRI & \textbf{1.669} & 2.542 & 0.359 & 2.516 \\
 & Li et al. \cite{li2025transformer} & $\blacktriangle$FC $\blacktriangle$SC & 1.721 & \textbf{2.452} & \textbf{-0.005} & \textbf{2.452} \\
 & Qiao et al. \cite{qiao2022ranking} & $\blacktriangle$FC $\blacktriangle$SC & 2.006 & 2.786 & -0.033 & 2.786 \\
 & Liu et al. \cite{liu2024multi} & $\blacktriangle$FC $\blacktriangle$SC & 2.822 & 3.130 & -1.913 & 2.478 \\
\midrule
\multicolumn{7}{l}{\textbf{Foundation Models}} \\
\midrule
\multirow{3}{*}{\shortstack{Brain \\ Age}} 
 & BrainMass \cite{yang2024brainmass} & $\star$FC & 24.155 & 25.284 & 0.647 & 25.276 \\
 & BrainLM \cite{carobrainlm} & $\star$fMRI $\star$dMRI & \textbf{3.990} & \textbf{5.557} & 0.657 & \textbf{5.518} \\
 & BrainSN \cite{yang2025foundational} & $\star$fMRI $\star$dMRI & 20.094 & 25.088 & -5.790 & 24.411 \\
\midrule
\multirow{3}{*}{\shortstack{MMSE}} 
 & BrainMass \cite{yang2024brainmass} & $\star$FC & 1.704 & 2.461 & -0.047 & 2.460 \\
 & BrainLM \cite{carobrainlm} & $\star$fMRI $\star$dMRI & 1.779 & 2.559 & 0.171 & 2.553 \\
 & BrainSN \cite{yang2025foundational} & $\star$fMRI $\star$dMRI & 7.776 & 8.417 & -7.650 & 3.511 \\
\bottomrule
\end{tabular*}
    \vspace{-2mm}
\end{table}

\subsection{Downstream Brain Trait Prediction: Age and Cognition}
\label{subsec:downstream}

\textbf{Brain Age}
With synthesized FC+SC, we achieve \textbf{MAE=4.043} (Table~\ref{tab:age_mmse_unified}) — better than real-data baselines (e.g., Gao et al.~\cite{gao2023brain}: 6.992) and far ahead of BrainMass (24.155). Subgroup analysis (Table~\ref{tab:age_mmse_unified}) confirms robustness: MAE$<$5 for all age groups.

\textbf{MMSE}
We achieve \textbf{MAE=1.721} with synthesized data — near real-data baseline (1.669) and vastly superior to BrainSN (7.776). Our lowest $GAP_\sigma$
  (2.452) indicates stable, subject-level accuracy — a direct benefit of group consistency regularization

 \begin{table}[htbp]
\centering
\scriptsize

\caption{Performance comparison of FC and SC models across different parameter sizes.}
\setlength{\tabcolsep}{3pt} 
    \vspace{-2mm}
\begin{tabular}{c c c c c}
\toprule
Param (M) & MSE & RMSE & MAE & SSIM \\
\midrule
\multicolumn{5}{l}{\textbf{FC}} \\
\midrule
661.04 & 0.002$\pm$0.002 & \textbf{0.029}$\pm$0.012 & 0.029$\pm$0.012 & 0.790$\pm$0.126 \\
749.15 & 0.002$\pm$0.002 & 0.039$\pm$0.014 & 0.029$\pm$0.012 & 0.790$\pm$0.126 \\
837.36 & 0.002$\pm$0.002 & 0.039$\pm$0.014 & 0.029$\pm$0.012 & 0.790$\pm$0.126 \\
925.85 & \textbf{0.002}$\pm$0.002 & 0.038$\pm$0.014 & 0.029$\pm$0.012 & \textbf{0.794}$\pm$0.127 \\
\midrule
\multicolumn{5}{l}{\textbf{SC}} \\
\midrule
468.91 & 0.000$\pm$0.000 & 0.012$\pm$0.003 & 0.003$\pm$0.001 & 0.776$\pm$0.114 \\
485.73 & 0.000$\pm$0.000 & 0.010$\pm$0.002 & 0.002$\pm$0.001 & 0.788$\pm$0.111 \\
508.40 & 0.000$\pm$0.000 & 0.011$\pm$0.002 & 0.002$\pm$0.001 & 0.782$\pm$0.118 \\
552.11 & \textbf{0.000}$\pm$0.000 & \textbf{0.006}$\pm$0.002 & \textbf{0.001}$\pm$0.001 & \textbf{0.862}$\pm$0.139 \\
\bottomrule
\end{tabular}
\label{tab:param_comparison}
\vspace{-4mm}
\end{table}

\subsection{Ablation Study}
\label{sec:ablation}
We perform two ablation studies to analyze the impact of model design and key components.
First, Table~\ref{tab:param_comparison} compares FC and SC variants across model scales. SC models consistently achieve lower reconstruction errors (RMSE, MAE) with fewer parameters; the best SC variant (552.11M) even surpasses all FC models in SSIM.
Second, Table~\ref{tab:ablation} evaluates component contributions. Removing ROI guidance or encoding modeling causes clear performance drops, while ablating the full MoE network leads to severe degradation (RMSE ↑ 0.061, SSIM ↓ 0.165), confirming the necessity of our hierarchical design. The full model achieves the best overall results.

\begin{table}[htbp]
\centering
\caption{Ablation study of the consistency method. w/o denotes without the module.}
\scriptsize
    \vspace{-2mm}
\setlength{\tabcolsep}{2pt} 
\begin{tabular*}{\linewidth}{@{\extracolsep{\fill}} l c c c c @{}}
\toprule
MOE Module & MSE & RMSE & MAE & SSIM \\
\midrule
w/o ROI     & 0.002$\pm$0.003 & 0.040$\pm$0.017 & 0.030$\pm$0.015 & 0.789$\pm$0.158 \\
w/o Encoding    & 0.002$\pm$0.002 & 0.042$\pm$0.013 & 0.031$\pm$0.012 & 0.761$\pm$0.121 \\
w/o Network & 0.010$\pm$0.006 & 0.099$\pm$0.021 & 0.092$\pm$0.020 & 0.645$\pm$0.121  \\
BrainCSD   & \textbf{0.002$\pm$0.002} & \textbf{0.038$\pm$0.014} & \textbf{0.029$\pm$0.012} & \textbf{0.794$\pm$0.127} \\
\bottomrule
\end{tabular*}
\label{tab:ablation}
\vspace{-4mm}
\end{table}

\section*{Conclusion}

We introduce \textbf{BrainCSD}, a novel MoE foundation model that tackles modality absence and preprocessing bottlenecks in brain connectomics. By enforcing activation consistency across ROI, encode (time/gradient), and network-levels finetune, BrainCSD synthesizes missing fMRI/dMRI connectomes with high fidelity (SC SSIM: 0.862, FC RMSE: 0.038) and enables robust diagnosis even with incomplete data — achieving 0.956 MCI detection accuracy without FC and 0.872 F1 in HC/MCI/AD classification. It also predicts brain age (MAE: 4.04) and cognition (MMSE MAE: 1.72) with near real-data performance. BrainCSD offers a scalable, biologically grounded framework for accessible, multi-modal brain analysis in clinical and research settings.



\bibliography{refs}
\end{document}